\documentclass[reprint,
superscriptaddress,amsmath,amssymb,aps, longbibliography
]{revtex4-2}

\usepackage[utf8]{inputenc}

\usepackage[lining,semibold]{libertine} 
\usepackage{amsthm}
\usepackage[libertine, cmintegrals, bigdelims, vvarbb]{newtxmath}

\usepackage{amsmath}
\usepackage{amsfonts}
\usepackage{mathrsfs}
\usepackage{gensymb}
\usepackage{bbm}
\usepackage{dsfont}

\usepackage{kbordermatrix}
\renewcommand{\kbldelim}{(}
\renewcommand{\kbrdelim}{)}

\usepackage{chemformula}
\usepackage[caption=false]{subfig}

\usepackage{soul}
\usepackage{xcolor}

\theoremstyle{definition}

\usepackage{scalerel} 

\definecolor{webgreen}{rgb}{0,.5,0}
\definecolor{webbrown}{rgb}{.6,0,0}
\definecolor{grigio}{rgb}{.85,.85,.85} 
\definecolor{RoyalBlue}{rgb}{0.0, 0.14, 0.4}
\definecolor{skyblue1}{rgb}{0.45,0.62,0.81}
\definecolor{skyblue2}{rgb}{0.2,0.39,0.64}
\definecolor{skyblue3}{rgb}{0.13,0.29,0.53}
\definecolor{scarlet1}{rgb}{0.93,0.16,0.16}
\definecolor{scarlet2}{rgb}{0.8,0,0}
\definecolor{scarlet3}{rgb}{0.64,0,0}

\definecolor{g}{gray}{0.50}

\usepackage{hyperref}
\hypersetup{%
    colorlinks=true, linktocpage=true, pdfstartpage=1, pdfstartview=FitV,%
    breaklinks=true, pdfpagemode=UseNone, pageanchor=true, pdfpagemode=UseOutlines,%
    plainpages=false, bookmarksnumbered, bookmarksopen=true, bookmarksopenlevel=1,%
    hypertexnames=true, pdfhighlight=/O,
    urlcolor=webbrown, linkcolor=RoyalBlue, citecolor=webgreen, 
    pdftitle={},%
    pdfauthor={Francesco Avanzini},%
    pdfsubject={},%
    pdfkeywords={},%
    pdfcreator={pdfLaTeX},%
    pdfproducer={LaTeX REVTeX}%
}

\newcommand{\pos}{\boldsymbol r}

\newcommand{\chemspecies}{\alpha}

\newcommand{\labelspecies}{Z}

\newcommand{\elrct}{\rho}

\newcommand{\curr}{ j_\elrct }

\newcommand{\flux}[1]{ j_{#1\elrct} }

\newcommand{\stcoeff}[1]{\nu^{\chemspecies}_{{#1\elrct}}}

\newcommand{\RNum}[1]{\uppercase\expandafter{\romannumeral #1\relax}}

\usepackage{graphicx}
\usepackage{dcolumn}
\usepackage{bm}
\usepackage{siunitx}
\usepackage[noend]{algorithmic}
\usepackage{algorithm}
\usepackage[skins]{tcolorbox}
\usepackage{setspace}
\usepackage{blkarray}

\usepackage{xcolor}
\usepackage{soul}
\usepackage{systeme}
\usepackage[capitalize]{cleveref}

\usepackage{scalerel} 

\makeatletter
\def\maketag@@@#1{\hbox{\m@th\normalfont\normalsize#1}}
\makeatother

\newcommand{\me}{\mathrm{e}}
\newcommand{\blue}[1]{\textcolor{black}{#1}}

\begin{document}
\title{Nonideal Reaction-Diffusion Systems: Multiple Routes to Instability}
\newcommand\unilu{\affiliation{Department of Physics and Materials Science, University of Luxembourg, L-1511 Luxembourg City, Luxembourg}}
\newcommand\unipdchim{\affiliation{Department of Chemical Sciences, University of Padova, Via F. Marzolo, 1, I-35131 Padova, Italy}}

\author{Timur Aslyamov}
\email{timur.aslyamov@uni.lu}
\unilu

\author{Francesco Avanzini}
\email{francesco.avanzini@unipd.it}
\unilu
\unipdchim

\author{\'Etienne Fodor}
\email{etienne.fodor@uni.lu}
\unilu

\author{Massimiliano Esposito}
\email{massimiliano.esposito@uni.lu}
\unilu



\begin{abstract}
We develop a general classification of the nature of the instabilities yielding spatial organization in open nonideal reaction-diffusion systems{, based on linear stability analysis}. This encompasses dynamics where chemical species diffuse, interact with each other, and undergo chemical reactions driven out-of-equilibrium by external chemostats. 
We find analytically that these instabilities can be of two types: instabilities caused by \blue{intermolecular energetic interactions (E-type),} and instabilities caused by multimolecular out-of-equilibrium chemical reactions \blue{(R-type)}.
Furthermore, we identify a class of chemical reaction networks, containing unimolecular networks \blue{but also extending beyond them}, that can only undergo \blue{E-type} instabilities. 
We illustrate our analytical findings with numerical simulations on two reaction-diffusion models, each displaying one of the two types of instability { and generating stable patterns}.
\end{abstract}
\maketitle


\textit{Introduction.}---Reaction-diffusion (RD) systems play a crucial role in explaining the emergence of many spatial structures across scales, e.g., spiral form of galaxies \cite{cross2009pattern}, predator-prey distributions in ecological models~\cite{segel1972dissipative}, skin color patterns of animals \cite{kondo2010reaction}, self-organization at the molecular scale~\cite{mikhailov2017chemical}, phase separation in electrochemical batteries~\cite{bazant2017thermodynamic}. The foundation of RD theory dates back to the seminal paper of A. M. Turing~\cite{turing1952}, where he proposed a spatial symmetry-breaking mechanism yielding stationary patterns.

Subsequent studies by the Brussels school of thermodynamics, led by I. Prigogine, showed the physicochemical and thermodynamical relevance of Turing patterns. Since they considered ideal solutions where the concentration dynamics is governed by linear diffusion and mass-action kinetics, they emphasized the need to consider multimolecular reactions and open systems driven far from equilibrium to generate patterns~\cite{Lefever,Prigogine1971,Nicolis1977}. Indeed, in ideal solutions, on the one hand, mutimolecular reactions are necessary to generate purely entropic interactions between species which create the nonlinearities at the basis of the instabilities, and on the other hand, nonequilibrium drives are required to prevent relaxation towards homogeneous concentration profiles. 
Since then, RD structures in ideal solutions have been extensively studied~\cite{marcon2016high, diego2018key, haas2021turing, brauns2020phase}.

nonideal mixtures feature instead both entropic and energetic interactions, so that concentrations can be non-homogeneous at equilibrium even in absence of chemical reactions. This is well described, for instance, by the Cahn-Hilliard theory of spinodal decomposition~\cite{cahn1958free}. Recently, nonideal mixtures that undergo chemical reactions have attracted considerable attention due to their role in biology~\cite{weber2019physics}. 
Reactions can affect the nature of phase separation and, when driven out of equilibrium, these active systems exhibit rich phenomenologies. 
To be physicochemically justified and thermodynamically consistent, RD models need to express not only diffusion but also chemical dynamics in terms of nonideal chemical potentials \cite{lefever1995comment,carati1997chemical}.
Heuristic models of nonideal diffusion reactions have been considered, but use ideal chemical kinetics (mass action) and thus lack thermodynamic consistency~\cite{glotzer1995reaction, Cates2010,lutsko2016mechanism}. 
Consistent models have been considered in Refs.~\cite{weber2019physics, kirschbaum2021controlling, zwicker2022intertwined, bauermann2022energy}. 
However, they focus on unimolecular reactions, which cannot accommodate any spatial instability in the absence of energetic interactions. 
Extending these studies by considering multimolecular reactions is important because the instabilities that cause spatial organization can now arise from an interplay between chemical reactions and molecular interactions.

In this Letter, we consider thermodynamically consistent \blue{deterministic} descriptions of generic nonideal mixtures of species undergoing diffusion and chemical reactions of arbitrary molecularity, driven out of equilibrium by external chemostats. 
Using linear stability analysis, we provide a rigorous classification of the possible instabilities and predict the conditions under which they arise.
We find that that they can be of two distinct types, which we call \blue{{\em E-type} and {\em R-type}} instabilities. The former depends solely on the details of \blue{the intermolecular interactions, as in Cahn-Hilliard theory of spinodal decomposition}. The latter is controlled by the topology of the chemical reaction networks (CRNs), \blue{as in Turing theory of instabilities in ideal mixtures.} 
\blue{Unlike other classifications of RD-instabilities \cite{Cross1993a,frohoff2023nonreciprocal} which focus solely on dynamics, ours is based on the underlying microscopic mechanism causing the instability.}
Importantly, we identify a wide class of CRNs where the instability can only be of \blue{E-type}. We also illustrate our findings with two specific models, each of them displaying one of two types of instabilities.

\textit{Chemical reactions and molecular interactions.}---We consider an isothermal nonideal mixture at temperature $T$, composed of chemical species~$\alpha\in\mathcal{S}$ which are reacting and diffusing within a solution of volume~$V$.  
We partition the set of chemical species $\mathcal{S}$ into two non-overlapping subsets: 
the internal species $x\in\mathcal{X}$ and the chemostated species $y\in\mathcal{Y}$. 
The latter are exchanged with the external chemostats. Each chemical reaction $\rho\in \mathcal{R}$ is represented by the chemical equation
\begin{equation}
\label{eq:reactions}
\nu^y_{+\rho} Z_y + \nu^x_{+\rho} Z_x \ch{ <=>[ $+\rho$ ][ $-\rho$ ]}
\nu^{y}_{-\rho} Z_{y} + \nu^{x}_{-\rho} Z_{x}\,,
\end{equation}
where $\labelspecies_\chemspecies$ is the chemical symbol of species $\alpha\in\mathcal S$, and $\stcoeff{+}$ (resp. $\stcoeff{-}$) is the stoichiometric coefficient of species $\chemspecies$ in the forward (resp. backward) reaction $+\rho$ (resp. $-\rho$). 
We always use Einstein notation: repeated upper-lower indices implies the summation over all the allowed values for the indices. 
The set of internal species may include a non-reacting species ($\text{nr}\in\mathcal X$), defined by  $\nu^{\text{nr}}_{\pm\rho}=0$ for all $\pm\rho$. 
We assume that the chemostatted species $\mathcal{Y}$ are ideal and maintained at constant homogeneous concentrations, which result in the homogeneous chemical potentials $\mu_y$. 
In practice, chemostats can drive chemical reactions far from equilibrium.

Turning to the dynamics, by combining dynamical density functional theory~\cite{te2020classical} and open CRNs theory~\cite{Rao2016, Avanzini2021a, Avanzini2022}, the concentration fields of internal species $c_x(\boldsymbol{r},t)$ 
evolve as
\begin{align}
    \label{eq:RD-def}
    \partial_tc_x &= D_x\boldsymbol{\nabla}\cdot(c_x\boldsymbol{\nabla}\mu_x) + S_x^{\rho}j_\rho\,, 
\end{align}
with closure relations
\begin{subequations}
    \label{eq:closures}
    \begin{align}
    \label{eq:cp-def}
    \mu_x &= \frac{1}{k_\text{B} T} \dfrac{\delta F}{\delta c_x}\,, 
    \\
    \label{eq:current-def}
     \curr & = \flux{+}- \flux{-}\,, 
     \\
     \label{eq:react-fluxes}
j_{\pm\rho}&=s_\rho\me^{\mu_x\nu^x_{\pm\rho}+\mu_y\nu^y_{\pm\rho}}\,,
    \end{align}
\end{subequations}
where $\boldsymbol{\nabla}$ is the spatial gradient; $k_\text{B}$ is the Boltzmann constant; 
$\mu_x$ (resp. $D_x$) is the non-dimensional chemical potential (diffusion coefficient) of species~$x$; 
$F$ is the Helmholtz free energy of the nonideal mixture;
$S^{\rho}_x = \nu_x^{-\rho} - \nu_x^{+\rho}$ is the entry of the so-called stoichiometric matrix $\mathbb{S}$ of the internal species (indexes $x$ and $\rho$ correspond to the rows and columns, respectively);
$\curr$ is the net current of reaction $\rho$
expressed as the difference between the forward $\flux{+}$ and the backward $\flux{-}$ reaction flux; $s_\rho$ is a positive preexponential factor that depends on the activation energy of the reaction $\rho$. The reaction fluxes $j_{\pm\rho}$ are defined in \cref{eq:react-fluxes} as the Arrhenius like rates. 
\blue{
We note that thermodynamically consistent currents could in principle allow for an additional dependence on the concentrations in $s_\rho(\boldsymbol{c})$, but is rarely considered and is thus omitted in our study. 
}
The diffusive contribution to the dynamics describes a pure gradient flow and the only nonequilibrium drive stems from chemostats. In absence of chemostats, the system relaxes to equilibrium. 
Dynamical models similar to \cref{eq:RD-def,eq:closures} have been recently considered in Refs.~\cite{kirschbaum2021controlling, zwicker2022intertwined, miangolarra2023non}.

The Helmholtz free energy of the nonideal mixture reads
\begin{equation}
     \label{eq:F_model}
        F[\boldsymbol{c}] =k_\text{B}T\int d\pos f(\boldsymbol{c}, \boldsymbol{\nabla}\boldsymbol{c})+F_\text{chm}\,,
\end{equation}
where $f$ is given in terms of the gradient expansion
\begin{equation}
\label{eq:f_model}
        f (\boldsymbol{c}, \boldsymbol{\nabla}\boldsymbol{c})= f_0(\boldsymbol{c})+\frac{1}{2} K_{x,x'}(\boldsymbol{c}) \big(\boldsymbol{\nabla}c^{x}\big)\cdot\big(\boldsymbol{\nabla}c^{x'}\big)\,, 
\end{equation}
the constant term $F_\text{chm}$ is the contribution due to the ideal chemostatted species,
and $\boldsymbol{c} = (c_1,\dots,c_{|\mathcal{X}|})^\intercal$ (with $|\mathcal{X}|$ being the number of internal species). 
Here, $f$, $f_0$, $K_{x,x'}=K_{x',x}$ for $x,x'\in \mathcal{X}$ are model functions of the concentrations. 
\Cref{eq:f_model} is consistent with the free energy used in Refs.~\cite{saha2020scalar,Joanny2020}, and it can be straightforwardly extended to a more general form, including higher orders in gradients, as introduced by Cahn and Hilliard~\cite{cahn1958free}.

By using Eqs.~\eqref{eq:F_model} and~\eqref{eq:cp-def}, the chemical potentials read
\begin{equation}
\mu_x = \frac{\partial f_0}{\partial c_x}+\frac{1}{2}\frac{\partial K_{x',x''}}{\partial c_x} (\boldsymbol{\nabla}c^{x'})\cdot(\boldsymbol{\nabla}c^{x''})-\boldsymbol{\nabla}\cdot(K_{x,x'}\boldsymbol{\nabla} c^{x'})\,.
\label{eq:cp_model}
\end{equation}
For ideal solutions, the free energy only comprises the entropic term $f=\sum_x c_x \ln c_x $, in which case one recovers a linear diffusion and mass-action kinetics in~\cref{eq:RD-def}: $D_x\boldsymbol{\nabla}\cdot(c_x\boldsymbol{\nabla}\mu_x) = D_x \boldsymbol{\nabla}^2 c_x$
and $j_{\pm\rho}\propto\prod_x c_x^{\nu^x_{\pm\rho}}$. For nonideal mixtures, the local free energy $f_0$ contains additional contributions, typically given as an expansion in powers of the concentrations~\cite{zwicker2022intertwined}, yielding non-linear diffusion. 
If one considers homogeneous concentrations 
$\boldsymbol{c}^*=(c_1^*,\dots,c_{|\mathcal{X}|}^*)^\intercal$, we have $\boldsymbol{\nabla}\mu_x^*=0$ where $\mu_x^*=\partial f_0(\boldsymbol{c}^*)/\partial c_x$. Therefore, a homogeneous fixed point of \cref{eq:RD-def} must satisfy the following steady-state condition: 
\begin{equation}
    \label{eq:fixed_point}
    S_x^\rho j_\rho(\{\mu_x^*\},\{\mu_y\}) = 0\,. 
\end{equation}
Equation~\eqref{eq:fixed_point} shows that fixed points are determined by the chemical reaction contribution to the dynamics, which depends on (i) the stoichiometric coefficients, (ii) the chemical potentials of chemostatted species $\mu_y$, and (iii) the details of the local free-energy $f_0$ (i.e., including both entropic and energetic contributions). In contrast, for purely diffusive systems, each concentration is conserved, so that the homogeneous concentrations are fixed independently of the free-energy parameters.

\textit{Nature of instabilities: \blue{E-type vs R-type}.---}To analyze the stability of the homogeneous steady state, we consider small concentration perturbations around the homogeneous fixed point $c_x(\pos,t) = c_x^* + \delta c_x(\pos, t)$.
Using the Fourier transform $\Tilde{g}(\boldsymbol{q})=\int d\pos g(\pos)\exp{(\text{i} \boldsymbol{q}\cdot\pos)}$, the perturbation of the chemical potentials in \cref{eq:cp_model} can be written as:
\begin{subequations}
\begin{align}
        \delta\Tilde{\boldsymbol{\mu}}(q) &= \mathbb{M}(q)\cdot\delta\Tilde{\boldsymbol{c}}(q)\,,
\\
 \label{eq:matM}
        M_{xx'}(q) &=  \frac{\partial^2 f_0(\boldsymbol{c}^*)}{\partial c_{x} \partial c_{x'}}+q^2K_{x,x'}(\boldsymbol{c}^*)\,,
\end{align}    
\end{subequations}
where 
$\delta\Tilde{\boldsymbol{\mu}}=(\delta\Tilde{\mu}_1,\dots,\delta\Tilde{\mu}_{|\mathcal{X}|})^\intercal$ and 
$\delta\Tilde{\boldsymbol{c}}=(\delta\Tilde{c}_1,\dots,\delta\Tilde{c}_{|\mathcal{X}|})^\intercal $. 

Using \cref{eq:RD-def}, the evolution of a perturbation $\delta\Tilde{\boldsymbol{c}}$ reads:
\begin{subequations}
\label{eq:rate_equation_linear_fourier}
\begin{align}
         \partial_t \delta\Tilde{\boldsymbol{c}}(q)&=-q^2\mathbb{A}\cdot\delta\Tilde{\boldsymbol{\mu}}(q)+
        \mathbb{S}
         \cdot
          \delta\Tilde{\boldsymbol{j}}(q) 
         \,,\\
    \mathbb{A} &=\text{diag}\big(D^{}_1c_1^*,\dots, D^{}_{|\mathcal{X}|}c_{|\mathcal{X}|}^*\big)\,.    
\end{align}
\end{subequations}
By inserting $\mu_x=\mu_x^*+\delta\mu_x$ into \cref{eq:react-fluxes} and calculating the Fourier transform, we arrive at
\begin{equation}\small
    \delta\Tilde{j}_{\rho} (q) =s_{\rho} \,\Big[ \nu^{x}_{+\rho}\, \me^{\mu_{x'}^*\nu^{x'}_{+\rho}+\mu_y\nu^y_{+\rho}} - \nu^{x}_{-\rho} \,\me^{\mu_{x'}^*\nu^{x'}_{-\rho}+\mu_y\nu^y_{-\rho}} \Big] \,\delta\Tilde{\mu}_{x} (q)\,.
    \label{eq:perturbation_currents}
\end{equation}
Using \cref{eq:matM}, we deduce that \cref{eq:rate_equation_linear_fourier} becomes
\begin{subequations}
\begin{align}
    \label{eq:rate-linear-Fourier}
    \partial_t \delta\Tilde{\boldsymbol{c}}(q) &=\mathbb{B}(q)\cdot\mathbb{M}(q)\cdot\delta\Tilde{\boldsymbol{c}}(q)\,,\\
     \label{eq:matB}
     \mathbb{B}(q)&=-q^2\mathbb{A}+\mathbb{C}\,.
\end{align}    
\end{subequations}
The elements of the square matrix $\mathbb{C}$ are defined from \cref{eq:perturbation_currents}:
\begin{equation}
    \label{eq:matC}
    C_{x}^{x'}=S_x^\rho s_\rho \,\Big[ \nu^{x'}_{+\rho}\me^{\mu_{x''}^*\nu^{x''}_{+\rho}+\mu_y\nu^y_{+\rho}} - \nu^{x'}_{-\rho} \,\me^{\mu_{x''}^*\nu^{x''}_{-\rho}+\mu_y\nu^y_{-\rho}}\Big]\,,
\end{equation}
where $x$ and $x'$ are the row and column index, respectively. 
\blue{
We emphasize that the product structure of the Jacobian matrix $\mathbb{B}(q) \cdot \mathbb{M}(q)$ follows from the thermodynamically consistent description defining both diffusion and chemical fluxes in terms of chemical potentials.}

Standard stability analysis~\cite{cross2009pattern} of
\cref{eq:rate-linear-Fourier} implies that the homogeneous fixed point $\boldsymbol{c}^*$ is unstable if at least one of the eigenvalues $\{\lambda_i\}$ of the \blue{Jacobian} matrix $\mathbb{B}(q)\cdot\mathbb{M}(q)$ has a positive real part for a given wavenumber $q$. 
To avoid any divergence of the perturbations $\delta\boldsymbol{\tilde c}(q)$ at small wavelengths, we impose that all eigenvalues $\lambda_i$ are negative as $q$ tends to infinity \blue{\cite{cross2009pattern}}. In practice, this can be enforced by choosing appropriately $\{K_{x,x'}\}$ in \cref{eq:f_model} which determines the cost of forming interfaces.
\blue{This means that if we assume that $\text{Im}\,{\lambda_i}(q_0)=0$, the condition for the homogeneous fixed point $\boldsymbol{c}^*$ to become unstable, for at least one wavenumber $q_0\neq0$, can be expressed in terms of the determinant of the Jacobian matrix:}
\begin{equation}
    \label{eq:instability_condition}
    \det \big(\mathbb{B}(q_0)  \cdot \mathbb{M}(q_0)\big) = \big(\det\mathbb{B}(q_0) \big) \,\big(\det\mathbb{M}(q_0)\big) = 0\,.
\end{equation}
\blue{
Our analysis covers instabilities  which typically induce stationary or transient patterns \cite{Cross1993a}.
However, it does not cover instabilities often arising in homogeneous time-oscillations and traveling waves \cite{Cross1993a}, where simultaneous $\text{Re}\,{\lambda_i}(q_0)=0$ and $\text{Im}\,{\lambda_i}(q_0)\neq 0$. In that case, the instability condition can not be expressed in terms of the determinant of the matrix}
{

The condition in \cref{eq:instability_condition} shows that the instability can be caused by two distinct mechanisms: $\det\mathbb{M}(q)=0$ or $\det\mathbb{B}(q)=0$.  
Matrix $\mathbb{M}$ depends on the free energy~\eqref{eq:F_model}, and it also characterizes the purely diffusive system without reactions ($\mathbb{C}=0$). The case $\det\mathbb{M}(q)=0$ can only happen due to energetic interactions, since for ideal mixture $\mathbb{M}$ is diagonal and positive. Thus, we refer to such an instability as E-type. In contrast, $\det \mathbb{B}=0$ can happen in either ideal or nonideal solutions. The corresponding instability is not caused by energetic interactions, but instead by multimolecular chemical reactions. We refer to it as an R-type instability. Although matrix $\mathbb{C}$ contains information on both the free energy (via the chemical potentials) and the stoichiometric matrix $\mathbb S$, the condition $\det \mathbb{B}=0$ can only be met if $\mathbb S$ satisfies certain conditions independent of the free energy, as we discuss below.

} 

\textit{Restricted route to instability.}---We now identify the specific class of CRNs where  only instabilities of E-type can arise. In this class, each reaction $\rho$ interconverts $m_\rho$ molecules of one specific internal species into $m_\rho$ molecules of a different internal species, without constraints on the stoichiometry of the chemostatted species:
\begin{equation}
\label{eq:cond}
\nu^y_{+\rho} Z_y + m_\rho \varepsilon_{x,+\rho} Z_{x} \ch{ <=>[ $+\rho$ ][ $-\rho$ ]}
\nu^{y}_{-\rho} Z_{y} + m_\rho \varepsilon_{x',-\rho} Z_{x'}\,,
\end{equation}
where $x\neq x'$, and there is no summation over $x$ and $x'$ (since they do not appear as repeated upper-lower indices). In~\cref{eq:cond}, $\varepsilon_{x,\pm\rho}$ and $\varepsilon_{x',\pm\rho}$ can be either $0$ or $1$, so that every internal species is either a reactant or a product in a given reaction $\rho$. Furthermore,
$m_\rho>0$ is an integer number that can be different for each reaction $\rho$. 

To prove that the CRNs (\ref{eq:cond}) can only undergo \blue{E-type} instabilities, we first demonstrate in \cref{sec:appendix-matC} that the corresponding $\mathbb{C}$ [\cref{eq:matC}] has non-negative non-diagonal elements 
\begin{equation}
\label{eq:matC-non-diagonal}
    C_{x,x'}\geq 0\,,\quad x\neq x'.
\end{equation}
and that the diagonal elements of $\mathbb C$ satisfy the inequality
\begin{equation}
    \label{eq:matC-diagonal}
   C_{x,x} \leq - \sum_{x'\neq x \in \mathcal{X}}C_{x',x}\,,
\end{equation}
where the equality holds if and only if $\sum_{x\in {\cal X}}S_x^\rho=0$, i.e., when the CRNs conserves the total concentration~\cite{Rao2016}. 
We then proceed to show that \cref{eq:matC-non-diagonal,eq:matC-diagonal} imply $\det \mathbb{B}(q)\neq0$, which rules out \blue{R-type} instability based on \cref{eq:instability_condition}. 
To this end, we note that every eigenvalue of the matrix $\mathbb{B}$ lies in the complex plane within (at least) one of a series of circles, referred to as Gershgorin circles~\cite{varga2010gervsgorin}. In practice, each Gershgorin circle has a radius $R_x$ defined as
    \begin{align}
    \label{eq:Gershgorin-radius}
         &R_x = \sum_{x'\neq x\in \mathcal{X}} |B_{x',x}| = \sum_{x'\neq x\in\mathcal{X}} C_{x',x}\,,
    \end{align}
where we have used that $\mathbb A$ is diagonal. Moreover, the center of each Gershgorin circle is located on the real axis (since $\mathbb B$ has only real elements) at the point $v_x$ given by
    \begin{align}
    \label{eq:Gershgorin-center}
         &v_x =B_{x,x} \leq  -q^2 D_x c_x^* - R_x \,,
    \end{align}
where we used~\cref{eq:matC-diagonal}. As one can see from \cref{eq:Gershgorin-radius,eq:Gershgorin-center}, for $q>0$ all Gershgorin circles are entirely located in the left complex half-plane. Thus, all eigenvalues of the matrix $\mathbb{B}$ have a negative real part, so that $ \det \mathbb{B}(q) \neq 0$. Combining this result with the condition in~\cref{eq:instability_condition}, it follows that the only way for the RD systems with reactions in~\cref{eq:cond} to entail any instability is $\det {\mathbb M}(q)=0$, namely via a \blue{E-type} instability. 

Reactions described by~\cref{eq:cond} include  pseudo-unimolecular ($\forall \rho$: $\nu^y_{\pm\rho} \geq 0$ and $m_\rho=1$) and non-unimolecular reactions (for at least one $\rho$: $m_\rho>1$). It is well known that ideal RD systems made of pseudo-unimolecular reactions cannot exhibit Turing patterns, as their dynamics is linear. In nonideal mixtures, energetic contributions to the free energy make the dynamics non-linear even for pseudo-unimolecular reactions. Indeed, the reaction fluxes in \cref{eq:react-fluxes} and
the matrix $\mathbb{B}$ in~\cref{eq:matB} explicitly depends on $\partial f(\boldsymbol{c}^*)/\partial c_x$ through the chemical potential $\mu_x$ [\cref{eq:cp_model}], which could a priori trigger \blue{R-type} instabilities. Yet, our result
shows that nonlinearities stemming from molecular interactions can only create \blue{E-type} instabilities for pseudo-unimolecular reactions. Importantly, this result extends to a special class of CRNs that also includes non-unimolecular reactions. We emphasize that, although such chemical reactions cannot generate \blue{R-type} instabilities, their topology and rates strongly influence the location of the \blue{E-type} instability, as they determine the homogeneous fixed points [\cref{eq:fixed_point}].

\begin{figure}[t]
    \centering
    \includegraphics[width=0.47\textwidth]{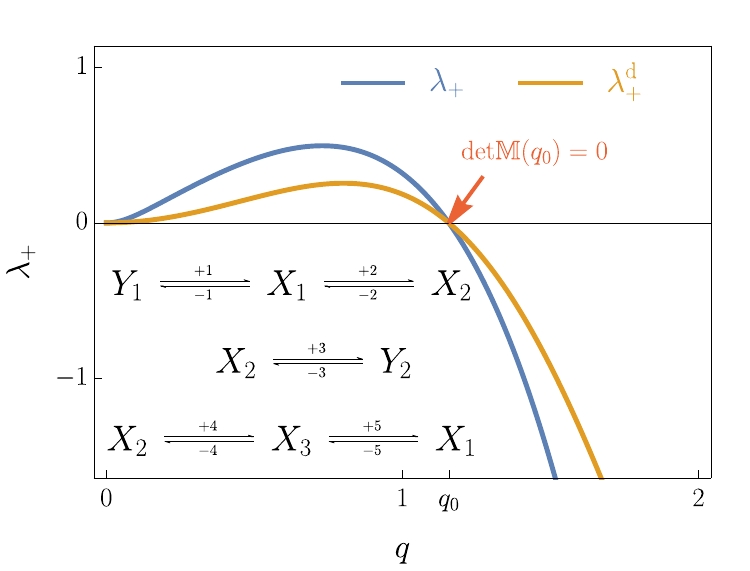}
    \vspace{-0.25cm}
    \caption{Inset: \blue{Example of CRN that can only undergo E-type instabilities}. Main: 
    The blue curve is $\lambda_+$ for the RD system from the inset, the orange curve is $\lambda^\text{d}_+$ for the corresponding purely diffusive system with same fixed points \blue{$\boldsymbol{c}^*=(0.12, 2.72, 0.10, 1.00)^\intercal$}. 
    Parameters:  $D_x=1$, $\mu_1^\theta=\mu_3^\theta=\mu_\text{nr}^\theta=0$, $\mu_2^\theta=-2$, $s_\rho=10^{-3}$, $\chi=1$, $k_1=0.5$, $k_2=0.1$, $\mu_{Y_1}=1$, $\mu_{Y_2}=-1$ in arbitrary units.
    }    
    \vspace{-0.25cm}
    \label{fig:eigenvalues_model-D-broken}
\end{figure}

\textit{Illustrative examples.}---We \blue{first consider} the pseudo-unimolecular CRN in the inset of \cref{fig:eigenvalues_model-D-broken}.  
In \cref{sec:appendix-example}, we derive its matrix $\mathbb{C}$ and show that it satisfies \cref{eq:matC-non-diagonal,eq:matC-diagonal}. Thus, this CRN belongs to the special class which admits only \blue{E-type} instabilities.
For the chemical potentials, we use the following expressions:
\begin{subequations}
\begin{align}
    \label{eq:example-cp}
        & \mu_x =\mu^\theta_x + \log c_x + L_{x,x'} c^{x'} -  K_{x,x'} {\boldsymbol\nabla}^2 c^{x'}\,,\\
\small
&\mathbb{L} = \begin{blockarray}{ccccccc}
 & & \color{gray} X_1 & \color{gray} X_2 & \color{gray}  X_3  &\color{gray} \text{nr}\\
\begin{block}{cc (ccccc)}
\color{gray} X_1 & & 0 & \chi & 0 & 0 \\
\color{gray} X_2 & & \chi & 0 & \chi & \chi \\
\color{gray} X_3 & & 0 & \chi & 0 & 0 \\
\color{gray} \text{nr} & & 0 & \chi & 0 & 0 \\
\end{block}
\end{blockarray}\,\,,\quad
\mathbb{K} =
\begin{blockarray}{ccccccc}
 & & \color{gray} X_1 & \color{gray} X_2 & \color{gray}  X_3 &  \color{gray} \text{nr}\\
\begin{block}{cc (ccccc)}
\color{gray} X_1 & & k_1 & k_2 & 0 & 0 \\
\color{gray} X_2 & & k_2 & k_1 & k_2 & k_2 \\
\color{gray} X_3 & & 0 & k_2 & k_1 & 0  \\
\color{gray} \text{nr} & & 0 & k_2 & 0 & k_1 \\
\end{block}
\end{blockarray}
\,\,,
\end{align}
\end{subequations}
\blue{where $\mu_x^\theta$ are the standard chemical potentials.}

The matrices $\mathbb{L}$ and $\mathbb{K}$ describe the mean-field molecular interactions and the cost at forming interfaces, respectively. We numerically determine the homogeneous fixed points $\boldsymbol{c}^*$, from which we compute the matrices $\mathbb{M}$ and $\mathbb{B}$ according to Eqs.~\eqref{eq:matM} and~\eqref{eq:matB}, respectively. 
In \cref{fig:eigenvalues_model-D-broken}, \blue{for one of the fixed points} we plot the eigenvalue $\lambda_+(q)$ of the matrix $\mathbb{B}\cdot\mathbb{M}$ which becomes positive over a finite range of $q$. We compared its behavior with the eigenvalue $\lambda_+^\text{d}(q)$ corresponding to pure diffusion system with $\mathbb{B}^\text{d}=-q^2 \mathbb{A}$ and $\mathbb{M}^\text{d}=\mathbb{M}$ at the same fixed point $\boldsymbol{c}^*$. Both $\lambda_+(q)$ and $\lambda_+^\text{d}(q)$ are positive in the same range of $q$, then vanish at the same point $q_0$ satisfying $\det \mathbb{M}(q_0)=0$. This agrees with the scenario of \blue{E-type} instabilities: it is sufficient to analyze the eigenvalues of the purely diffusive system to deduce the range of stability of the corresponding RD system. 
\blue{The model displays at least two additional homogeneous fixed points. One is stable $\lambda_+<0$. The other is unstable, and it is such that $\lambda_+(0)>0$ and $\lambda_+(q)$ reaches a maximum at $q_\text{m}>0$.}

\begin{figure}[t]
    \centering
    \includegraphics[width=0.47\textwidth]{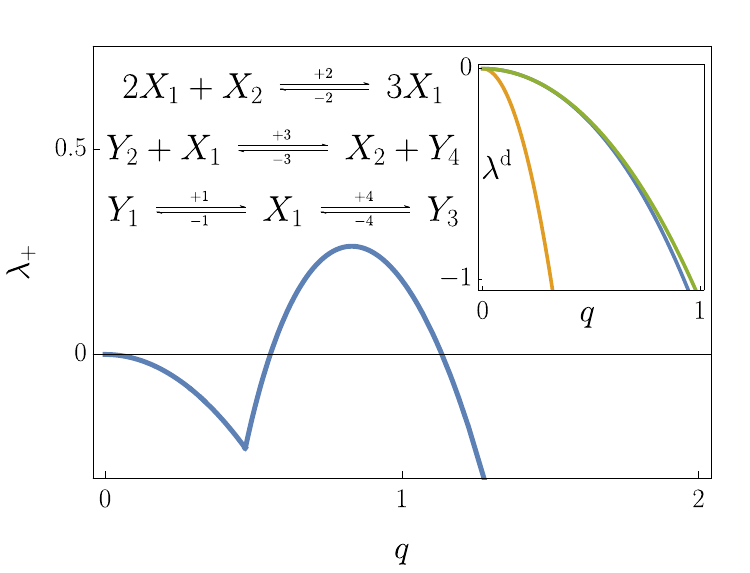}
    \vspace{-0.25cm}
    \caption{
    Inset (left): \blue{Example of CRN undergoing a R-type instability.}
    Main: The blue curve is $\lambda_+$. Inset (right): The eigenvalues of the corresponding purely diffusive dynamics. Parameters: $D_1=D_\text{nr}=1$, $D_2=10$, $\mu_1^\theta=\mu_\text{nr}^\theta=0$, $\mu_2^\theta=9.2$, $s_\rho=1$, $\chi=0$, $k_1=0.1$, $k_2=0.05$, $\mu_{Y_1}=9.9$, $\mu_{Y_2}=1.1$, $\mu_{Y_3}=-27.6$, $\mu_{Y_4}=-18.4$ in arbitrary units.
    }  
    \vspace{-0.25cm}
    \label{fig:eigenvalues_model-C-Brusselator}
\end{figure}

\begin{figure}
    \centering
    \includegraphics[width=0.5\textwidth]{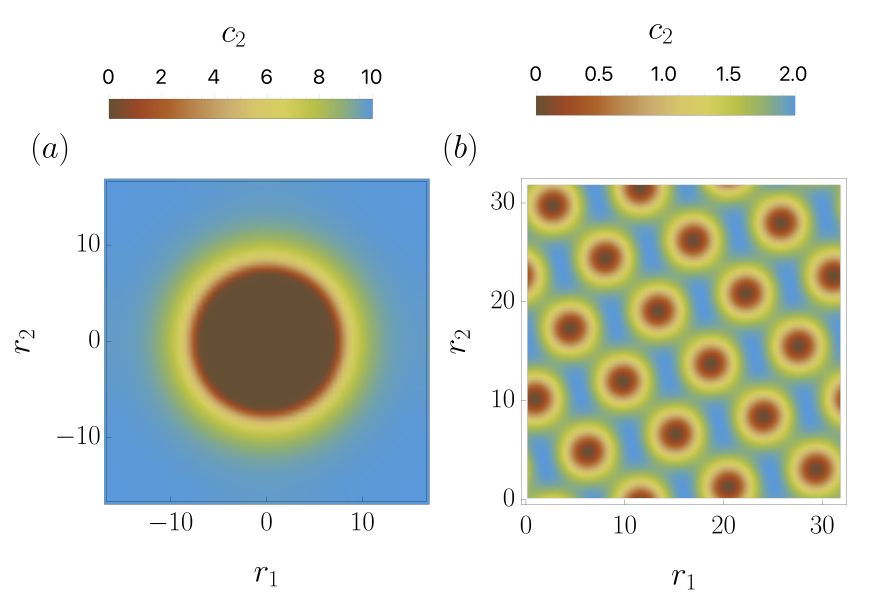}
    \vspace{-0.45cm}
    \caption{The steady-state patterns for $c_2$ from numerical simulations in two-dimensional space with the periodic boundary conditions. (a) The unimolecular CRN with parameters from \cref{fig:eigenvalues_model-D-broken}; (b) The Brusselator with parameters from \cref{fig:eigenvalues_model-C-Brusselator}. For simulations we used the py-pde package \cite{py-pde}.
    }
    \vspace{-0.25cm}
    \label{fig:patterns}
\end{figure}

We now \blue{consider a nonideal version of the} Brusselator model in the inset of \cref{fig:eigenvalues_model-C-Brusselator}. We use $\mu_1$, $\mu_2$ from \cref{eq:example-cp}.
This model does not satisfy the conditions in \cref{eq:cond} and thus \blue{can and does display an R-type} instability.  
The right inset in \cref{fig:eigenvalues_model-C-Brusselator} shows that while all eigenvalues of the purely diffusive process are negative, the eigenvalues of the RD Brusselator can be positive (\cref{fig:eigenvalues_model-C-Brusselator}), in stark contrast with the previous model. Indeed, since the Brusselator is not restricted to \blue{E-type} instabilities, it can exhibit patterns in a regime where the corresponding purely diffusive system is stable.

Finally, we compare the results of numerical simulations for the two RD systems in~\cref{fig:eigenvalues_model-D-broken,fig:eigenvalues_model-C-Brusselator}. \Cref{fig:patterns} shows the steady-state 2d patterns of one of the chemical components.
It is worth noting that the pattern in \cref{fig:patterns}(a), for the RD system with reactions satisfying \cref{eq:cond}, is qualitatively analogous to the complete phase separation obtained in purely diffusive dynamics. In contrast, the pattern of the Brusselator in \cref{fig:patterns}b shows a striking different spatial organization which could not be reproduced by the corresponding purely diffusive system. 

\textit{Discussion.}---We \blue{characterized the nature of instabilities in thermodynamically consistent deterministic} dynamics of nonideal RD \blue{systems in solution} { described by Arrhenius rates}.
We considered reversible reactions, but our results also hold for irreversible reactions.
{ 
Extension to mixtures \cite{weber2019physics} is left for future work.}
{ 
Our decomposition in either \blue{E-type} or \blue{R-type} instability was shown for generic free energies and chemical reactions obeying the Arrhenius rate law \cref{eq:react-fluxes}. It cannot be extended to rates laws 
where $s_\rho(\boldsymbol{c})$ depends on the concentration (due to the definition of the matrix $\mathbb{B}$). It can however be extended when $s_{\rho}(\boldsymbol{\mu})$ depends on the chemical potentials. Our proof that CRNs (\ref{eq:cond}) can only undergo \blue{E-type} instabilities however does not hold in this latter case and should be revisited.
}
Future studies will be needed to study the thermodynamics (resp. fluctuations) of these \blue{systems}, as recently done for ideal solutions~\cite{Falasco2018a,Avanzini2019a,Avanzini2020a} (resp. ideal \cite{falasco2023macroscopic}
and nonideal RD systems~\cite{tiani2023stochastic}),
and clarify their connection to heuristic active field theories including chemical degrees of freedom~\cite{Markovich2021, Cates2022}. 
{
Turing instability in ideal solution requires different diffusion coefficients \cite{haas2021turing}, but our condition for R-type instability does not. It would be interesting to explore if interactions could generate R-instabilities inducing stationary patterns in models with same diffusion coefficients.
}
\blue{Our framework provides a tool to analyze how interactions promote or suppress the different types of instabilities, as recently done for models without thermodynamic consistency\cite{menou2023physical,luo2023influence}.}

\textit{Acknowledgments}---This research was funded by project ChemComplex (C21/MS/16356329) and project SMAC (14389168) funded by FNR (Luxembourg), and by project INTER/FNRS/20/15074473 funded by F.R.S.-FNRS (Belgium) and FNR (Luxembourg).


\appendix
\section{Proof of (\ref{eq:matC-non-diagonal}) and (\ref{eq:matC-diagonal}) for CRNs that satisfy (\ref{eq:cond})}
\label{sec:appendix-matC}

First, we recognize that the stoichiometric coefficients of the chemical equations~\eqref{eq:cond} satisfy the following constraints: 
\begin{subequations}
\label{eq:special_class-S-appendix}
\begin{align}
\label{eq:special_class-S-cond_0}
&\text{if}\quad \nu_{x,\pm\rho} \neq 0\quad \text{then}\quad \nu_{x,\mp\rho} = 0 \,,\\
\label{eq:special_class-S-cond_1}
&\text{either}\quad\nu_{x,\pm\rho} = 0
\quad\text{or}\quad\nu_{x,\pm\rho} = m_\rho
\,,\\
\label{eq:special_class-S-cond_2}
&\text{either}\quad\sum_{x} \nu_{x,\pm\rho} = 0
\quad\text{or}\quad\sum_{x} \nu_{x,\pm\rho} = m_\rho
\,.
\end{align}
\end{subequations}

Second, we split the set of chemical reactions $\mathcal{R}$ into two groups: internal and exchange reactions $\mathcal{R}=\mathcal{R}_\text{in}\cup\mathcal{R}_\text{ex}$.
Internal reactions $\rho \in \mathcal{R}_\text{in}$ conserve the total concentration, i.e., $\sum_{x} S_{x,\rho}=0$ for $\rho\in\mathcal{R}_\text{in}$. 
The exchange reactions do not, i.e., $\sum_{x} S_{x,\rho}\neq0$ for $\rho\in\mathcal{R}_\text{ex}$, and read
\begin{align}
\label{eq:ex_reactions}
\nu^y_{+\rho_\text{ex}} Z_y&\ch{ <=>[ $+\rho_\text{ex}$ ][ $-\rho_\text{ex}$ ]}
\nu^{y'}_{-\rho_\text{ex}} Z_{y'} + \nu_{x,-\rho_\text{ex}} Z_{x}\,,
\end{align}
namely, the internal species $x$ is either a reactant or a product.


Third, we consider the non-diagonal elements $C_{x,x'}$ with $x' \neq x$.
i)~Because of \cref{eq:special_class-S-cond_2}, $S_{x,\rho}=\pm m_{\rho}$ implies $\nu_{x',\mp\rho}=0$. 
ii)~Because of \cref{eq:ex_reactions}, 
$S_{x,\rho_\text{ex}}=\pm m_{\rho_\text{ex}}$ implies $\nu_{x',\rho_\text{ex}} = \nu_{x',-\rho_\text{ex}}=0$.
Hence, the terms of the summation over $\rho$ in \cref{eq:matC}, namely, 
\begin{equation}
    \Xi_{x,x',\rho} = S_{x,\rho} s_\rho\Big[ \nu_{x',+\rho}\me^{\mu_{x''}^*\nu^{x''}_{+\rho}+\mu_y\nu^y_{+\rho}} - \nu_{x',-\rho}\me^{\mu_{x''}^*\nu^{x''}_{-\rho}+\mu_y\nu^y_{-\rho}}\Big]\,,
    \label{eq:refinC}
\end{equation}
read
\begin{subequations}
\begin{align}
         & \Xi_{x,x',\rho} =m_\rho s_{\rho}e^{\mu_\alpha^*\nu^{\alpha}_{\pm\rho}}\nu_{x',\pm\rho}\geq 0\,,\quad\text{if}\quad S_{x,\rho}=\pm m_\rho\,,    \\
         &\Xi_{x,x',\rho} =0 \,,\quad\text{if}\quad S_{x,\rho}=0\,,\\
         &\Xi_{x,x',\rho} =0 \,,\quad\text{if} \quad\rho\in\mathcal{R}_{\text{ex}}\,,\label{eq:offdiagEX}
\end{align}
\end{subequations}
showing that all non-diagonal elements in \cref{eq:matC} are non-negative, and thus proving \cref{eq:matC-non-diagonal}.


Fourth, we consider the diagonal elements $C_{x,x}$.
Because of Eqs.~\eqref{eq:special_class-S-cond_0} and~\eqref{eq:special_class-S-cond_1}, 
$S_{x,\rho}=\pm m_\rho$ implies $\nu_{x,\mp\rho}=m_\rho$ and $\nu_{x,\pm\rho}=0$.
Hence,  the terms of the summation over $\rho$ in \cref{eq:matC} become
\begin{equation}
\Xi_{x,x,\rho} =- m_\rho s_{\rho}e^{\mu_\alpha^*\nu^{\alpha}_{\mp\rho}}\nu_{x,\mp\rho}\leq 0\,,\quad\text{if}\quad S_{x,\rho}=\pm m_\rho\,,
\end{equation}
independently of whether $\rho\in\mathcal{R}_{\text{in}}$ or $\rho\in\mathcal{R}_{\text{ex}}$.


Fifth, we consider $\sum_x C_{x,x'}$.
By using the splitting $\mathcal{R}=\mathcal{R}_\text{in}\cup\mathcal{R}_\text{ex}$,
Eq.~\eqref{eq:offdiagEX},
and $\sum_{x} S_{x,\rho}=0$ for $\rho\in\mathcal{R}_\text{in}$, we obtain 
\begin{equation}
   \sum_x C_{x,x'}  =
   \sum_{\rho\in\mathcal{R}_{\text{ex}}} \Xi_{x',x',\rho} \leq  0\,,
\end{equation}
proving \cref{eq:matC-diagonal}.

\section{Matrix $\mathbb{C}$ for \cref{fig:eigenvalues_model-D-broken}}
\label{sec:appendix-example}
Here we derive the matrix $\mathbb{C}$ for the example shown in \cref{fig:eigenvalues_model-D-broken}. 
The coefficients $\nu_{\alpha,\pm\rho}$ can be written in matrix form:
\begin{align}
\label{eq:example-nu}
\{\nu_{\alpha,+\rho}\}=
\begin{blockarray}{ccccccc}
 & & \color{gray} 1 & \color{gray}2 \color{gray}& \color{gray}3 & \color{gray}4 & \color{gray}5 \\
\begin{block}{cc (ccccc)}
\color{gray} X_1 & & 0 & 1 & 0 & 0 & 0 \\
\color{gray} X_2 & & 0 & 0 & 1 & 1 & 0 \\
\color{gray} X_3 & & 0 & 0 & 0 & 0 & 1 \\
\color{gray} Y_1 & & 1 & 0 & 0 & 0 & 0 \\
\color{gray} Y_2 & & 0 & 0 & 0 & 0 & 0 \\
\end{block}
\end{blockarray}\,\,,\,
\{\nu_{\alpha,-\rho}\}=
    \begin{blockarray}{ccccccc}
 & & \color{gray} 1 & \color{gray}2 \color{gray}& \color{gray}3 & \color{gray}4 & \color{gray}5 \\
\begin{block}{cc (ccccc)}
\color{gray} X_1 & & 1 & 0 & 0 & 0 & 1 \\
\color{gray} X_2 & & 0 & 1 & 0 & 0 & 0 \\
\color{gray} X_3 & & 0 & 0 & 0 & 1 & 0 \\
\color{gray} Y_1 & & 0 & 0 & 0 & 0 & 0 \\
\color{gray} Y_2 & & 0 & 0 & 1 & 0 & 0 \\
\end{block}
\end{blockarray}\,\,.
\end{align}
which gives us the stoichiometric matrix $\mathbb{S}$
\begin{equation}
\label{eq:example-matS}
\mathbb{S}=
    \begin{blockarray}{ccccccc}
 & & \color{gray} 1 & \color{gray}2 \color{gray}& \color{gray}3 & \color{gray}4 & \color{gray}5 \\
\begin{block}{cc (ccccc)}
\color{gray} X_1 & & 1 & -1 & 0 & 0 & 1 \\
\color{gray} X_2 & & 0 & 1 & -1 & -1 & 0 \\
\color{gray} X_3 & & 0 & 0 & 0 & 1 & -1 \\
\color{gray} \text{nr} & & 0 & 0 & 0 & 0 & 0 \\
\end{block}
\end{blockarray}\,\,.
\end{equation}
Using \cref{eq:example-matS,eq:example-nu,eq:matC}, for \cref{fig:eigenvalues_model-D-broken} we find that $\mathbb{C}=$
\begin{equation}
\footnotesize
\label{eq:example-matC}
\begin{blockarray}{cc cccc}
 & & \color{gray} X_1 & \color{gray} X_2 \color{gray}& \color{gray}X_3 & \color{gray}\text{nr} \\
\begin{block}{cc (cccc)}
\color{gray} X_1 & & -s_1\me^{\mu_1^*}-s_2\me^{\mu_1^*} - s_5\me^{\mu_1^*} & s_2\me^{\mu_2^*} & s_5\me^{\mu_3^*} & 0\\
\color{gray} X_2 & & s_2\me^{\mu_1^*} & -s_2\me^{\mu_2^*}-s_3\me^{\mu_2^*}-s_4\me^{\mu_2^*}& s_4\me^{\mu_3^*} & 0 \\
\color{gray} X_3 & & s_5\me^{\mu_1^*} & s_4\me^{\mu_2^*} & -s_4\me^{\mu_3^*}-s_5\me^{\mu_3^*} & 0   \\
\color{gray} \text{nr} & & 0 & 0 & 0 & 0 \\
 \end{block}
\end{blockarray}\,\,,
\end{equation}
where the fixed point chemical potential $\mu_i^*$ for $i=1,2,3$ correspond to the species $X_1$, $X_2$, $X_3$, respectively. From \cref{eq:example-matC} one can see that $\mathbb{C}$ satisfies the properties in \cref{eq:matC-non-diagonal,eq:matC-diagonal}. Notice that, as discussed in \cref{sec:appendix-matC}, the exchange reactions---rates $s_1$ and $s_3$---contribute only to the diagonal elements.



\bibliography{biblio}
\end{document}